\documentclass[reprint,amsmath,amssymb,aps,prd,superscriptaddress,preprintnumbers]{revtex4-2}
\usepackage{graphicx}
\usepackage{dcolumn}
\usepackage{bm}
\usepackage{xcolor}
\usepackage{bm}
\usepackage{url}     
\usepackage[utf8]{inputenc}
\usepackage{amsthm, amsfonts}
\usepackage{overpic}
\usepackage{float}
\usepackage[colorlinks=true,
            linkcolor=blue,
            citecolor=blue,
            urlcolor=blue]{hyperref}
\newcommand{\abs}[1]{\left|#1\right|}
\newcommand{\black}[1]{\mathbf{#1}}
\newcommand{\ket}[1]{|#1\rangle}
\newcommand{\bra}[2]{\langle#1|}
\graphicspath{{images/}}  

\begin{document}
 
\title{Vacuum entanglement in a time-dependent electric field}
 
\author{\'{A}lvaro \'{A}lvarez-Dom\'{\i}nguez}
\email{a.alvarez.dominguez@ucm.es}
\affiliation{Departamento de Econom\'{i}a Financiera y Actuarial y Estad\'{i}stica, \\ Universidad Complutense de Madrid, 
Plaza de las Ciencias 1, 
28040 Madrid, Spain}

\author{Luis J. Garay}
\email{luisj.garay@ucm.es}
\affiliation{Departamento de Física Te\'{o}rica and IPARCOS, \\ Universidad Complutense de Madrid, 
Plaza de las Ciencias 1, 
28040 Madrid, Spain}

\author{Mercedes Martín-Benito}
\email{m.martin.benito@ucm.es}
\affiliation{Departamento de Física Te\'{o}rica and IPARCOS, \\ Universidad Complutense de Madrid, 
Plaza de las Ciencias 1, 
28040 Madrid, Spain}

\author{Iker Sanz-González}
\email{iker.sanz@unizar.es}
\affiliation{Departamento de Física Te\'{o}rica, \\ Universidad de Zaragoza, Pedro Cerbuna 12, 
50009 Zaragoza, Spain}


\begin{abstract}

We analyze the correlations of the vacuum state of a charged scalar field by constructing localized observables in two disjoint regions. We consider a time-dependent homogeneous electric field and examine the time evolution of correlations by relating it with the particle-antiparticle creation through the Schwinger effect. Our results show that the electric field increases the mixedness of the states and enhances total correlations, while the distillable entanglement among the regions decreases, vanishing at a finite time for sufficiently intense pulses within the non-perturbative pair-production regime. We also discuss the effect of anisotropy induced by the external background, showing that the distribution of correlations acquires a nontrivial directional dependence.
 
\end{abstract}
\maketitle


\section{Introduction}\label{sec:introduction}

Correlations constitute a powerful tool for characterizing complex physical systems, as they are capable of quantifying the amount of shared information among different degrees of freedom. The first systematic attempt to quantify them was made by Shannon \cite{Shannon1948}, who introduced entropy as a measure of the uncertainty associated with a physical system. The advent of Quantum Mechanics (QM) lead to an extension of this framework in a profound way. In this context, the question of whether QM fundamentally described nature or, instead, there was an underlying local hidden-variable theory became a central philosophical issue \cite{EPR}. However, Bell derived a set of inequalities that any theory satisfying local realism must obey \cite{Bell1964}. Subsequent experiments showed unequivocally that these inequalities are violated in nature (see, e.g.,~\cite{Aspect1982}), thereby demonstrating that the universe follows the laws compatible with QM. Moreover, these developments revealed that, besides classical correlations, composite quantum systems may exhibit a genuinely quantum form of correlation with no classical counterpart: entanglement.

A composite quantum system is said to be entangled if it cannot be fully described in terms of the states of the corresponding individual constituents~\cite{Schrodinger1935}. This quantum correlation was first understood as a mere mathematical peculiarity of QM~\cite{EPR,Bell1964}. It was later recognized as a useful resource since it enables tasks that are impossible within classical physics. This aim of exploiting entanglement as a valuable tool has led to the development of a recent area of physics: quantum information (QI)~\cite{NielsenChunag}. QI includes protocols such as quantum teleportation or quantum cryptography and contributes to the emergence of new technologies, whose common basis is entanglement.

Entanglement, originally characteristic of QM, naturally arises in Quantum Field Theory (QFT), which is built from the combination of both QM and special relativity. For instance, currently the Standard Model of particle physics is a QFT that most accurately explains nature and its components. Hence, it is reasonable to study how correlations (classical and quantum) are distributed within spacetime. Nonetheless, there are two important differences when considering a QFT framework. First, a reduced state of a QFT is generally mixed (see \cite{HollandsSanders}) and, therefore, classical and quantum correlations may be difficult to distinguish. Second, the transition from~QM to~QFT entails the passage from a finite number of degrees of freedom to infinitely many of them, which in turn are defined in the whole spacetime. Although the entanglement structure is harder to characterize due to this reason, its structure is much richer than in   QM.

In particular,
entanglement is observed in the vacuum state of a QFT \cite{SummersWerner1985,SummersWerner1987}. This feature is grounded in the Reeh-Schlieder theorem \cite{ReehSchlieder1961}: the vacuum is a highly nonlocal state in the sense that local operators acting on any open region can generate a dense set of states in the Hilbert space. That is, one can generate any state to arbitrary accuracy by acting with an appropriate choice of operators, even if they are supported within an arbitrarily small region of spacetime. Therefore, correlations and specifically entanglement are ubiquitous in QFT \cite{HollandsSanders}. 
Over the past two decades, these foundational insights have evolved into the active field of Relativistic Quantum Information (RQI), which bridges QI science, QFT, and general relativity (see \cite{PeresTerno2004,MartinMartinez2011, MannRalph2012} for reviews). 
Pioneering works demonstrated that non-inertial motion and spacetime geometry fundamental to general relativity profoundly alter the perception, degradation, and distribution of entanglement across reference frames \cite{Alsing2003, FuentesSchullerMann2005, Ball2006}. A central paradigm within RQI is entanglement harvesting, initiated in \cite{Valentini1991,Reznik2003, Reznik2005}, whereby initially uncorrelated localized quantum systems can extract correlations, including entanglement, from the quantum field vacuum through local interactions with the field, even when the detector interaction regions are spacelike separated (see e.g. \cite{PozasKerstjensMartinMartinez2015}). RQI studies have extended these operational tools to a plethora of scenarios, including curved spacetimes such as black holes or cosmologies  \cite{MannRalph2012, VerSteegMenicucci2009, MartinMartinezMenicucci2012}, quantum energy teleportation and energy extraction
\cite{Hotta2008,MartinMartinez2013}, or relativistic quantum technologies \cite{Bruschi2014,Agnesi2018,Sidhu2021}, while also serving as a probe for gravitational effects \cite{Howl2018, PercheRagulaMartinMartinez2023}.

In this work, we focus on the study of the distribution of correlations, which is strongly dependent on the system under consideration \cite{CalabreseCardy2004, Eisert2010,ubiquitousQFT}. Nontrivial backgrounds can affect how entanglement is spread among the fields' degrees of freedom \cite{MartinMartinezMenicucci2014,AlvesCamilo2019}. Additionally, studying entanglement in cosmological spacetimes (see, e.g.,~\mbox{\cite{inflationPatricia, RibesMetidieri2026}}) is particularly interesting, since quantum correlations play a key role in the early universe during inflation \cite{MukhanovChibisov1981}. Here, we consider an external agent as the nontrivial background: a classical, spatially homogeneous, and time-dependent electric field. Specifically, the aim is to study the Schwinger effect---the spontaneous production of particle-antiparticle pairs from the vacuum caused by a strong external electric field---through the lens of correlations (see \cite{Ebadi2014,Kranas2023} for related studies). Although it was first formulated in the context of Quantum Electrodynamics  \cite{Schwinger1951}, the connection with cosmological particle production in expanding universes \cite{Parker1968, Parker1969, Parker1971} (e.g., during inflation) is readily apparent. Nevertheless, an important difference should be emphasized. While cosmological particle production is usually studied in isotropic backgrounds, such as in FLRW cosmologies, the electric field is fixed in a single direction, thereby introducing an intrinsic anisotropy into the system. Specifically, we will consider an electric pulse coupled to a quantum charged scalar field and study correlations between modes of the field supported on two space-like disjoint regions, paying special attention to their behavior as time evolves. 

In general, studying the distribution of correlations across the degrees of freedom of a field is not straightforward. This motivates the introduction of a systematic procedure to characterize such systems. Gaussian states, whose description is fully determined by their first and second statistical moments, provide a suitable setting for this purpose. Beyond their practical advantages, Gaussian states arise in a variety of physically relevant scenarios, such as quantum optics, ground and thermal states of quadratic Hamiltonians or coherent and squeezed states~\cite{Olivares2012}. Moreover, they are especially significant in continuous-variable QI, underpinning recent studies of entanglement harvesting~\cite{Trevison2018, Hackl2019, OsawaNambuYoshimoto2026}, multimode systems~\cite{Yamaguchi2020a, Yamaguchi2020b, Agullo2025}, and partner modes in QFT~\mbox{\cite{KlcoSavage2021a,KlcoSavage2021b,KlcoBeckSavage2023,GaoKlco2024,agullo2025partner1,agullo2025partner2, RibesMetidieri2025}}.

The structure of this work is as follows. Section \ref{sec:formalism} is divided into two parts: on the one hand, we review the dynamics of a charged scalar field minimally coupled to a general homogeneous and time-dependent electric field, and quantize it through the canonical approach; on the other hand, we present QI techniques used in QM and particularize some correlation measures---entropy, mutual information and logarithmic negativity---to Gaussian states, which we will consider in subsequent sections. In Section \ref{sec:setup} we consider a specific setup for a~\mbox{$(1+d)$}-dimensional Minkowski spacetime in which we show how to apply QI tools in QFT by defining local observables for a charged scalar field in localized spatial regions. Section \ref{sec:electric_1d} particularizes the previous setup to a $(1+1)$-dimensional flat spacetime with a Sauter-type electric pulse, allowing us to study the time evolution of the correlation measures. In Section \ref{sec:electric_3d} we analyze the case of a $1+3$ spacetime, focusing on the anisotropy produced by the electric field. Finally, we elaborate our conclusions in Section \ref{sec:conclusions}.

\textit{Notation}. We set $c= \hbar=1$, and use the signature~\mbox{$(-, +, \cdots, +)$}. The Fourier convention is 
\begin{align}
    \tilde{f}( \black{k})=\int \frac{\text{d}^d \black{x}}{(2\pi)^{d/2}} \ f(\black{x}) e^{-i \black{k} \cdot \black{x}}.
\end{align}

\section{Formalism}\label{sec:formalism}

\subsection{Canonical quantization}\label{Sec:can_quantization}

The dynamics of a massive complex scalar field $\Phi$ in a $(1+d)$-dimensional Minkowski spacetime, minimally coupled to an electromagnetic field $A_\mu$, is described by the Lagrangian 
\begin{align}
\mathcal{L}=-(D_\mu \Phi)^* (D^\mu \Phi) - m^2 \Phi^* \Phi,
\end{align}
where \mbox{$D_\mu =\partial_\mu + i q A_\mu$} is the covariant derivative, $m$ is the mass and $q$ the charge of the field. The corresponding Euler-Lagrange equation yields the Klein-Gordon (KG) equation:
\begin{align}
(D_\mu D^\mu - m^2)\Phi=0.
\end{align}

The vector space of all smooth solutions of this equation is called the covariant phase space. The canonical phase space is the set of pairs~$(\Phi(\mathbf{x},t), \Pi(\mathbf{x},t))$, where~$\Pi=\partial \mathcal{L}/\partial (\partial_t \Phi)$  is the canonical momentum, thereby fully characterizing the classical system at a given time. The only non-vanishing equal-time Poisson bracket is given by
\begin{align}
    \{\Phi(t,\mathbf{x}),\Pi(t,\mathbf{y})\}=\delta(\mathbf{x}-\mathbf{y}).
\end{align}
 The conjugate field,~$\Phi^*$, is governed by the complex conjugate of the KG equation, i.e., for a field of charge~$-q$. Moreover, the mixed Poisson brackets vanish.

Let us consider an electric field $\mathbf{E}(t)$ which is time-dependent and spatially homogeneous. A convenient choice for the potential that respects homogeneity is the temporal gauge, $A_\mu=(0,\mathbf{A}(t))$, so that~$\mathbf{E}(t)=-\dot{\mathbf{A}}(t)$.

One can solve the KG equation in Fourier space. Due to the spatial homogeneity of the problem, solutions are of the form $\varphi_\mathbf{k}(t) e^{i \mathbf{kx}}$, where all the temporal dependence is encoded in the modes $\varphi_\black{k}(t)$. They satisfy a decoupled system of oscillator-like equations of the form
\begin{equation}
    \ddot{\varphi}_\mathbf{k}(t)+\omega^2_\mathbf{k}(t) \varphi_\mathbf{k}(t)=0,
    \label{eq:harmonic}
\end{equation}
with time-dependent frequency
\begin{equation}
    \omega_\mathbf{k}(t) =\sqrt{\mathbf{k}^2+2q \abs{\black{k}} \abs{\black{A}(t)} \cos{\theta} +q^2 \abs{\mathbf{A}(t)}^2+m^2},
    \label{eq:freq}
\end{equation}
which differs from the free frequency $\sqrt{\mathbf{k}^2 + m^2}$ not only in the time-dependence but also in the emergence of anisotropy caused by the angle $\theta$ between the electric field $\mathbf{E}(t)$ and the wave vector $\mathbf{k}$. 

Since the KG equation is of second order, we choose a suitable basis of solutions: the `particle' sector given by $\varphi_\mathbf{k}(t) e^{i \mathbf{kx}}$  and the `antiparticle' one, which is~\mbox{$\varphi_{-\mathbf{k}}^*(t) e^{-i \mathbf{kx}}$}, where $\varphi_\mathbf{k}(t)$ is any complex solution to Eq.~\eqref{eq:harmonic} normalized according to the time-independent Wronskian,
\begin{align}
\varphi_\mathbf{k}(t) \dot{\varphi}^*_\mathbf{k}(t)-\varphi^*_\mathbf{k} (t)\dot{\varphi}_\mathbf{k}(t)=i.
\end{align}
As a result, the general solution to this KG equation is
\begin{align}
 \!\!\!   \Phi(t, \mathbf{x})\!=\!\int \!\frac{\text{d}^d \mathbf{k}}{(2 \pi)^{d/2}} \!\left[a_\mathbf{k} \varphi_\mathbf{k}(t) e^{i \mathbf{kx}} + b^*_{-\mathbf{k}} \varphi_{-\mathbf{k}}^*(t) e^{-i \mathbf{kx}} \right]\!.
    \label{eq:sol_KG}
\end{align}
The  coefficients $a_\black{k}$ and $b_\black{k}$ fulfill the non-vanishing Poisson brackets
\begin{align}
    \{a_\black{k}, a_\black{q}^*\}=\{b_\black{k}, b_\black{q}^*\}=-i \delta(\black{k}-\black{q}),
\end{align}
as can be seen from the fact that $\Phi$ and $\Pi$ are canonical.

In order to quantize the classical theory, one can make use of the canonical quantization formalism, in which the expansion coefficients $a_\mathbf{k}$ and $b^*_\mathbf{k}$ are promoted to annihilation and creation operators $\hat{a}_\mathbf{k}$ and $\hat{b}_\mathbf{k}^\dagger$, respectively. They satisfy the nonzero commutation relations
\begin{equation}
    [\hat{a}_\black{k}, \hat{a}^\dagger_\black{q}]=[\hat{b}_\black{k}, \hat{b}^\dagger_\black{q}]=\delta(\black{k} - \black{q}),
    \label{eq:comm_creation}
\end{equation}
which in turn imply that $\hat{\Phi}$ and $\hat{\Pi}$ are indeed canonical. These operators define a vacuum state~\mbox{$\ket{0}$ such that $\hat{a}_\black{k} \ket{0}= \hat{b}_\black{k} \ket{0}=0$} for all $\black{k}$, upon which the repeated action of the corresponding creation operators generate a basis of the so-called Fock space. It should be noted that the notion of vacuum is not uniquely determined as it depends on the choice of mode decomposition in Eq. \eqref{eq:sol_KG}, i.e., on the specific choice of $\varphi_\black{k}(t)$. For a general review of canonical quantization in nontrivial backgrounds, see~\cite{BirrellDavies1982}.

\subsection{Quantum Information in QM}\label{sec: QI}

To examine correlation measures in QM, let us consider a bipartite system $A$ and $B$ whose respective Hilbert spaces are $\mathcal{H}_A$ and $\mathcal{H}_B$. The states of the joint system~$AB$ can be described by the tensor product Hilbert space~\mbox{$\mathcal{H}=\mathcal{H}_A \otimes \mathcal{H}_B$}. Every quantum state can be represented as a density matrix $\hat{\rho}$, which encodes whether such a state is pure or mixed: in the former case, $\hat{\rho}^2 =\hat{\rho}$, whereas in the latter, $\hat{\rho}=\sum_i p_i \hat{\rho}_i$,  with $p_i$ the classical probability of the system being in the state $\hat{\rho}_i$.
Entanglement is generally understood as the lack of separability. For the considered bipartition, a general mixed state $\hat{\rho}$ is separable if and only if
\begin{align}
    \hat{\rho}= \sum_i p_i \hat{\rho}^{(i)}_A \otimes \hat{\rho}^{(i)}_B,
\end{align}
where $\hat{\rho}^{(i)}_{J}$ ($J=A,B$) denotes the reduced density matrix of subsystem $J$. 

We can now introduce three measures of correlation and discuss their suitability for quantifying entanglement and correlations in general.
    \paragraph*{\bfseries Von Neumann entropy.} It quantifies the degree of uncertainty associated with the system's state, defined as~\cite{NielsenChunag}
    \begin{align}
        S(\hat{\rho})=-\text{Tr}(\hat{\rho} \log_2 \hat{\rho}).
    \end{align} 
    It vanishes for pure states and reaches its maximum for maximally mixed ones. It is not, in general, a \emph{bona fide} entanglement measure. However, if the joint system $AB$ is pure ($S(\hat{\rho}) = 0$) but the reduced systems are not~($S(\hat{\rho}_A),\ S(\hat{\rho}_B)  > 0$), one can directly conclude that~$\hat{\rho}$ represents an entangled state, thus making it a faithful entanglement monotone for such cases. In QFT, the states under consideration are typically mixed and the von Neumann entropy does not constitute an entanglement measure.

    \paragraph*{\bfseries Mutual information.} It captures both classical and quantum correlations and is defined 
    as~\cite{NielsenChunag}
    \begin{align}
        \mathcal{I}(A,B) = S(\hat{\rho}_A)+S(\hat{\rho}_B) - S(\hat{\rho}).
    \end{align} 
    It is non-negative and vanishes if and only if the reduced states $A$ and $B$ are uncorrelated. Although it quantifies total correlations, it cannot   distinguish  classical correlations from genuinely quantum ones, i.e., entanglement.
    
    \paragraph*{\bfseries Logarithmic negativity.} It is based on the Positive Partial Transpose (PPT) criterion \cite{Peres1996,Horodecki1997}, which provides a practical method for detecting entanglement. A density matrix is a positive semi-definite operator and so is the partial transposition of the density matrix (i.e. transposition only on the $A$ or the $B$ sector) of any separable mixed state. Conversely, the PPT criterion establishes that, if this partial transposition results in a non-positive operator, this is a sufficient condition for the existence of entanglement. This motivates the introduction of the logarithmic negativity,
    \begin{align}
        E_\mathcal{N}(\hat{\rho})= \log_2 \| \hat{\rho}^{T_A} \|_1,
    \end{align}
    where $\| \hat{O}\|_1=\text{Tr}\sqrt{\hat{O}^\dagger \hat{O}}$ for an arbitrary operator $\hat{O}$ and~$\hat{\rho}^{T_A}$ denotes the partial transpose with respect to subsystem $A$. If $\hat{\rho}^{T_A}$ remains a positive semidefinite operator, $E_\mathcal{N}=0$ since $\text{Tr}(\hat{\rho}^{T_A})=1$. Conversely, if $\hat{\rho}^{T_A}$ has any negative eigenvalue, $E_\mathcal{N} >0$, thus signaling the presence of entanglement. Despite non-positivity being a sufficient condition for the presence of entanglement, it is not, in general, necessary. Put differently, a state can be entangled even if $\hat{\rho}^{T_A}$ is positive semi-definite. The PPT criterion captures only usable or distillable entanglement, while the so-called bound entanglement remains undetectable \cite{Horodecki1998BoundEntanglement,VidalWerner2002}. Consequently, only a non-vanishing logarithmic negativity represents a valid entanglement quantifier. However, according to Peres-Horodecki's theorem \cite{PeresHorodecki1996}, this condition becomes both sufficient and necessary in~$\mathcal{H}= \mathbb{C}^2 \otimes \mathbb{C}^2$ or~$\mathcal{H}=\mathbb{C}^2 \otimes \mathbb{C}^3$.

Having introduced these correlation measures, it is useful to provide a systematic procedure in order to compute them. Henceforth, closely following \cite{HacklBianchi2021}, we consider a $2N_A + 2N_B$ set of bosonic degrees of freedom in phase space written in a Darboux basis
\begin{align}
\hat{r}=(\hat{q}_A^1, \hat{p}_A^1, \dots, \hat{q}_B^1, \hat{p}_B^1, \dots),
\end{align}
whose components fulfill the canonical commutation relations~\mbox{$[\hat{r}^\alpha, \hat{r}^\beta]=i \Omega^{\alpha \beta}\ \hat{\mathbb{I}}$}, with $\Omega^{\alpha \beta}$ the components of the inverse of the symplectic structure $\Omega$, that in Darboux coordinates is given by
\begin{equation}
    \Omega= \bigoplus_{i=1}^{N_A+N_B} \begin{pmatrix}
        0 & -1 \\ 1 & 0
    \end{pmatrix}.
    \label{eq:symplectic_matrix}
\end{equation}
Each quadrature pair $(\hat{q}^i_J,\hat{p}^i_J)$, with $J=A,B$, represents a single-mode symplectic subsystem, i.e., a non-commuting pair that generates a Lie subalgebra independent from that generated by all the other pairs.

\subsection{Gaussian states}\label{sec: GS}
For simplicity,   we will consider Gaussian modes so that they are entirely characterized by their first and second statistical moments in a given state. These operators can always be redefined to have vanishing expectation values, and the second moments are given by
\begin{align}
    \sigma_{\alpha \beta}=\langle  [\hat{r}^\alpha, \hat{r}^\beta]_+ \rangle,
\end{align}
where $\sigma_{\alpha \beta}$ are the components of the covariance matrix~$\sigma$, which is real and symmetric, and $[ \cdot,\cdot ]_+$ denotes the anticommutator. For a bipartite system, the general structure of the covariance matrix is
\begin{equation}
    \sigma= \begin{pmatrix}
        \sigma_A & C \\ C^T & \sigma_B
    \end{pmatrix},
    \label{eq:covariance_matrix}
\end{equation}
where $\sigma_{A,B}$ are the covariance matrices of the reduced states and $C$  encodes the correlations between them. 

Within this framework, it is reasonable to compute the previously presented correlation measures through symplectically invariant quantities, since they should not depend on  symplectic transformations, i.e., change of coordinates that preserve the symplectic structure~\eqref{eq:symplectic_matrix}. These quantities are the so-called symplectic eigenvalues which are essentially the eigenvalues of the matrix~$i\Omega \sigma$. For a covariance matrix~$\sigma$ of dimension~\mbox{$2N \times 2N$} (where~$N$ may denote either~$N_J$ for the reduced states~$\sigma_J$ or~$N_A+N_B$ for the joint system), the symplectic eigenvalues appear in pairs~$\{ \pm \nu_1, \pm \nu_2, \dots, \pm \nu_N \}$, with~$\nu_i \geq 1$, for all $i$~\cite{Serafini2017}. The three measures introduced above can be written in terms of the symplectic eigenvalues.

\paragraph*{\bfseries Von Neumann entropy.} From its definition, one can write \cite{Serafini2017}
    \begin{equation}
   \!\!     S\!=\!\sum_{i=1}^N  \left[\frac{\nu_i\! +\! 1}{2}  \log_2 \!\left(\frac{\nu_i\!+\!1}{2} \right) -  \frac{\nu_i\!-\!1}{2}  \log_2 \!\left(\frac{\nu_i\!-\!1}{2} \right) \right]\!.
        \label{eq:entropy_symp}
    \end{equation}
    Note that $\nu_i=1$ does not contribute to the amount of entropy, corresponding to a pure mode. 

    \paragraph*{\bfseries Mutual information.} Since Eq. \eqref{eq:entropy_symp} is manifestly invariant under symplectic transformations, mutual information is trivially invariant as well.

    \paragraph*{\bfseries Logarithmic negativity.} Partially transposing $\hat{\rho}$ is equivalent to computing $\sigma^{T_A}$, given by
    \begin{equation}
        \sigma^{T_A}\!=\!T_A \sigma T_A,\quad T_A\!=\! \left(\bigoplus_{i=1}^{N_A} \begin{pmatrix}
            1 & 0 \\ 0 & -1
        \end{pmatrix}\! \right)\!\oplus\! \left(\bigoplus_{j=1}^{N_B} \mathbb{I}_{2 \times 2}\! \right)\!.
        \label{eq:partial_cov}
    \end{equation}
    It is important to note that, although they are equivalent, $T_A$ does not correspond to the same operation in the density matrix picture as in the covariance matrix formalism. Similarly, one can compute the symplectic eigenvalues of $\sigma^{T_A}$, which we denote by $\tilde{\nu}_i$, yielding the following expression for logarithmic negativity \cite{Serafini2017}: 
    \begin{equation}
        E_\mathcal{N}=\sum_{i=1}^{N_A+N_B} \max(0, -\log_2 \tilde{\nu}_i).
        \label{eq:logarithmic negativity_symp}
    \end{equation}
    A non-vanishing logarithmic negativity requires~\mbox{$0< \tilde{\nu}_i <1$} for some $i$, in contrast with the condition~\mbox{$\nu_i \geq 1$}, which must be satisfied by all physical states. In the density matrix picture, the condition~\mbox{$0< \tilde{\nu}_i <1$} is equivalent to the existence of negative eigenvalues of $\hat{\rho}^{T_A}$. Importantly, for Gaussian states, the PPT criterion can be extended to become a necessary and sufficient condition for $1 \times N$-mode bipartitions \cite{Simon2000}.

\section{Gaussian states for charged fields}\label{sec:setup}

In this section, we apply the tools developed in Sec.~\ref{sec:formalism} to a specific bipartite system. Consider a charged scalar field in a $(1+d)$-dimensional Minkowski spacetime in the vacuum state. Let $A$ and $B$ be two $d$-dimensional closed balls of radius $R$, centered at $\mathbf{x}_A$ and $\mathbf{x}_B$, respectively. In addition, we consider them as disjoint spacelike regions; consequently, the distance between their centers must satisfy $\abs{\mathbf{x}_A-\mathbf{x}_B} \geq 2R$.

\begin{figure} \centering
\includegraphics[width=0.95\columnwidth]{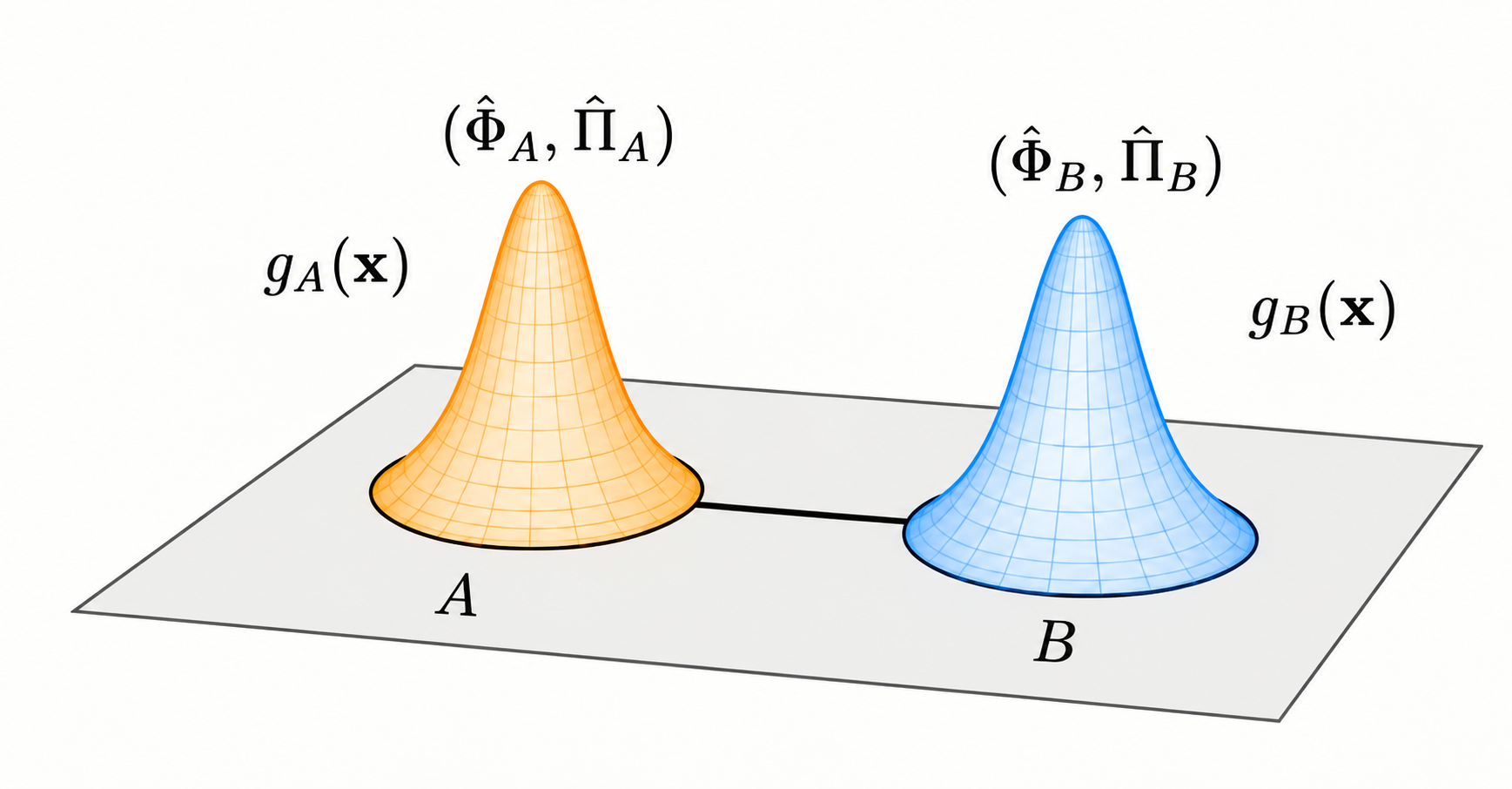}
	\caption{Illustration of two spatially separated spherical regions, with $g_{A,B}(\black{x})$ the smearing function defining the single modes $(\hat{\Phi}_{A,B}, \hat{\Pi}_{A,B})$.}
	\label{Fig:smearing_illustration}
\end{figure}

From the infinite number of degrees of freedom of the field we will select a finite number of them. For this restricted set, we can use the tools described in the previous section. We will choose the degrees of freedom by smearing. More explicit, let us define 
\begin{equation}
    \hat{\Phi}_J (t) = \int \text{d}^d \black{x}\ g(\black{x}-\black{x}_J) \hat{\Phi}(t,\black{x}).
    \label{eq:smeared_op}
\end{equation}
Here, $g(\black{x})$ is a smearing function with compact support within a ball of radius $R$ centred at the origin (see Fig.~\ref{Fig:smearing_illustration}). 
Therefore, $\hat{\Phi}_J$ is not a distribution-valued operator anymore but a localized well-defined operator, thus representing a single degree of freedom of the continuous field~$\hat{\Phi}(t,\black{x})$.

Within this setting, one can select $N_A$ and $N_B$ modes for each region respectively by drawing an analogy with the quadrature pairs $(\hat{q}_J,\hat{p}_J)$, as discussed in Sec. \ref{sec: QI}. An appropriate choice for a complex scalar field is
\begin{align}
    \hat{Q}_J^n  =  (\hat{\Phi}^n_J + \hat{\Phi}^{n\dagger }_J)/ {\sqrt{2}},\qquad 
   \hat{P}_J^n =   (\hat{\Pi}^n_J + \hat{\Pi}^{n \dagger }_J)/ {\sqrt{2}},
   \label{eq:modes}
\end{align}
with $n=1,\dots,N_J$, where  $\hat{\Phi}_J^n$ and $\hat{\Pi}_J^n$ denote the field and momentum smeared operators given by
\begin{align}
    \hat{\Phi}_J^n&= \int \text{d}^d \black{x}\  g^{(n)}(\black{x}-\black{x}_J) \hat{\Phi}(\black{x}),\\
    \hat{\Pi}_J^n&= K_n\int \text{d}^d \black{x}\ g^{(n)}(\black{x}-\black{x}_J) \hat{\Pi}(\black{x}).
\end{align}

Notice that we could have used a different smearing function for the field and its momentum. However, for simplicity, we choose the same one for both operators. In addition, the constant~$K_n$ is introduced so that~$\hat{\Phi}^n_J$ and~$\hat{\Pi}^n_J$ have the same physical dimensions. In fact, imposing the canonical commutation relations, \mbox{$[\hat{\Phi}^n_J, \hat{\Pi}^m_I]= i \delta^{nm} \delta_{JI}$}, uniquely fixes~\mbox{$K_n=\|g^{(n)}\|_2^{-2}$}, where~$\| \cdot \|_2$ is the $L^2$-norm.  

Regarding the smearing functions $\{g^{(n)}(\black{x}) \}$, they must satisfy certain requirements. First, their support is restricted to a ball of radius $R$ and their $L^2$-norm must be well-defined. Moreover, it is natural to demand spherical symmetry---the smearing function depends on~$\abs{\black{x}-\black{x}_J}$---due to the assumed geometry of the regions. Finally, each single-mode $(\hat{Q}^n_J, \hat{P}^n_J)$ must be algebraically independent from the others, as discussed in Sec. \ref{sec: QI}; that is, their commutators must vanish. This condition is trivially fulfilled for modes belonging to different regions, since their supports do not overlap. However, it becomes more subtle for modes defined within the same region. Therefore, one is led to consider a family of orthogonal functions such that
\begin{align}
\int \text{d}^d \black{x}\ g^{(n)}( \black{x}) g^{(m)} (\black{x}) = \delta^{nm}/{K_n}.
\end{align}
Taking these requirements into account, one can readily verify that the operators \eqref{eq:modes} define a suitable Darboux basis as they are Hermitian and satisfy the canonical commutation relations established by the symplectic structure $\Omega$. 

Note that, although $(\hat{\Phi}^n_J, \hat{\Pi}^n_J)$ are actually canonical operators and we could have taken them directly to form the Darboux basis, all the possible two-point functions among them vanish, which means that those modes are not correlated, whereas for a real scalar field this is not the case. Therefore, for a complex scalar field one needs to combine $(\hat{\Phi}^n_J, \hat{\Pi}^n_J)$ with $(\hat{\Phi}^{n \dagger}_J, \hat{\Pi}^{n \dagger}_J)$ as in Eq.~\eqref{eq:modes} in order to get non-trivial results.

Having introduced these elements, we can proceed to compute the anticommutators that define the components of the covariance matrix, which are given by
\begin{align*}
    \langle [\hat{Q}^m_I, \hat{Q}^n_J ]_+ \rangle  
    &= 2 \int \text{d}^d \black{k}\ \tilde{g}^{(m)}(\black{k}) \tilde{g}^{(n)}(\black{k}) \abs{\varphi_\black{k}}^2 \cos( \black{k} \cdot \bm{\ell}_{IJ}),
\end{align*}\vspace{-3ex}\begin{align*}
    \langle [\hat{P}^m_I, \hat{P}^n_J ]_+ \rangle 
    &= 2 \int \text{d}^d \black{k}\ \frac{\tilde{g}^{(m)}(\black{k}) \tilde{g}^{(n)}(\black{k})}{\| g^{(m)} \|_2^2\ \| g^{(n)} \|^2_2} \abs{\dot{\varphi}_\black{k}}^2 \cos( \black{k} \cdot \bm{\ell}_{IJ}),
 \end{align*}\vspace{-2ex}\begin{align}
 \langle [\hat{Q}^m_I, \hat{P}^n_J ]_+ \rangle 
    &=  \int \text{d}^d \black{k}\ \frac{\tilde{g}^{(m)}(\black{k}) \tilde{g}^{(n)}(\black{k})}{ \| g^{(n)} \|^2_2} \frac{d\abs{\varphi_\black{k}}^2}{dt} \cos( \black{k} \cdot \bm{\ell}_{IJ}),
    \label{eq:cross}
\end{align}
where $\bm{\ell}_{IJ}=\black{x}_I-\black{x}_J$ and $\tilde{g}^{(n)}(\black{k})$ denotes the Fourier transform of $g^{(n)}(\black{x})$.

It is worth emphasizing that the covariance matrix depends on time, since the modes $\varphi_\black{k}(t)$ are time-dependent due to the presence of an electric field. Consequently, the symplectic eigenvalues $\nu_j$ and $\tilde{\nu}_j$ also acquire a time dependence, which in turn determines the time evolution of the correlation measures between regions $A$ and $B$, regions that we consider fixed in time. 

Importantly, correlations are strongly subject to the choice of the smearing function or, equivalently, to the choice of modes. Hence, correlations are not solely a property of the state itself; they also depend on the particular chosen subsystem. 

\section{Correlations under electric fields in $1+1$ dimensions} \label{sec:electric_1d}

We now apply the previous formalism to $(1+1)$-dimensional Minkowski spacetime in the presence of a time-dependent electric field, $E(t)$. Studying correlations in $(1+1)$ dimensions constitutes a natural preliminary step towards the $(1+3)$-dimensional case. Correlations are typically stronger on lower spatial dimensions, anisotropic effects are trivially absent, and it is less computationally demanding. 

\subsection{Setup}
\begin{figure} \centering
\includegraphics[width=0.8\columnwidth]{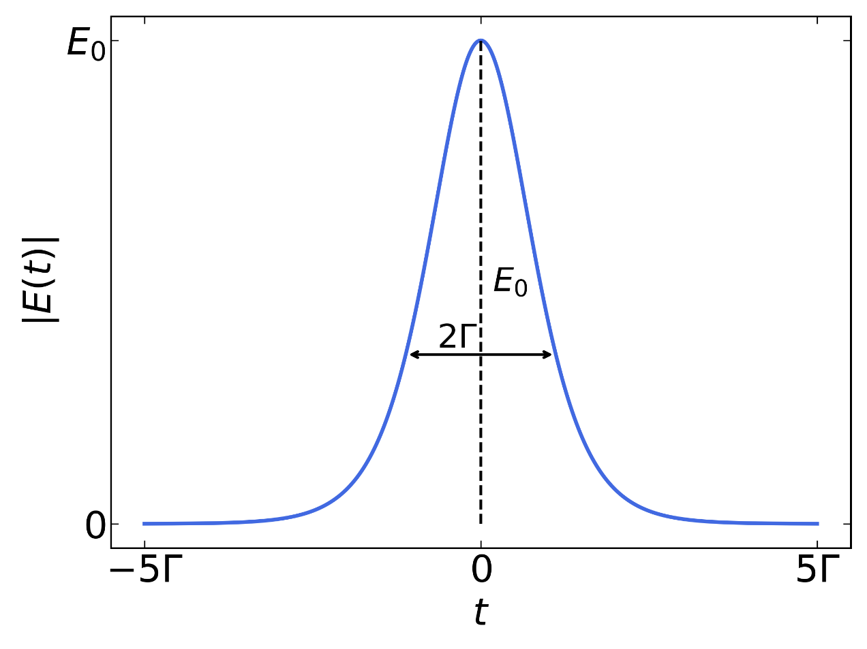}
	\caption{Sauter-type electric pulse with amplitude $E_0$ and a characteristic width parameter $\Gamma$, given by Eq.~\eqref{eq:electric_pulse}.}
	\label{Fig:electric_pulse}
\end{figure}

The purpose of this section is to understand how the correlation measures depend on the main characteristics of the electric field, and how they evolve over time. We consider an electric pulse that is switched on and subsequently switched off. In particular, let us study the case of the Sauter-type potential~\cite{Sauter1931},
\begin{equation}
\!\!\!    A(t)=E_0 \Gamma \left[\tanh{\left({t}/{\Gamma}\right)+1} \right],\quad E(t)= \frac{-E_0}{\cosh^2{(t /\Gamma)}},
     \label{eq:electric_pulse}
\end{equation}
which admits an exact analytical treatment. The pulse reaches its peak amplitude~$E_0$ at $t=0$ and vanishes asymptotically as $t\rightarrow \pm \infty$, with its effective width characterized by~$\Gamma$ (see Fig.~\ref{Fig:electric_pulse}).

To further specify our setup, we note that the scalar field mass~$m$ sets a characteristic scale through the inverse correlation length~\cite{ubiquitousQFT}. Although the dependence of entanglement on the mass is generally non-monotonic, working with relatively small masses allows us to capture a sizeable amount of correlations between the regions. Specifically, the regime $mR<1$ corresponds to a Compton wavelength larger than the size of the regions, thereby enhancing correlations between them. Similarly, increasing the separation between the two regions $\ell$ leads to further decrease of correlations, particularly entanglement, which is found to rapidly decay for $\ell >2R$ in all our simulations. Henceforth, we take $A$ and $B$ to be adjacent regions, $\ell =2R$. 

We must also specify the modes under consideration  in regions $A$ and $B$. To do so, we introduce the following family of orthonormal smearing functions:
\begin{equation}
    g^{(n)} (\black{x})=\sin{(2 \pi n r)} \Theta (1-r),\qquad \text{with } r=\abs{\black{x}}/R,
    \label{eq:senos}
\end{equation}
where $n=1, 2,3, \dots$ (we keep the vector notation although $\black{x} \in \mathbb{R}$). An advantage of this choice is the existence of a closed analytical expression for their Fourier transform, given by
\begin{equation}
    \tilde{g}^{(n)}(\black{k})=\frac{2 \sqrt{2 \pi} n R}{4 \pi^2 n^2-(kR)^2} \left[1-\cos{(kR)} \right],
    \label{eq:fourier_transform}
\end{equation}
where $k=\abs{\black{k}}$. An important property of this family of smearing functions is that $g^{(n)}(\black{x})$ becomes more oscillatory as $n$ increases, thus corresponding to modes with decreasing wavelength. In Fourier space, this is reflected in the fact that $\tilde{g}^{(n)}(\black{k})$ is localized around~\mbox{$kR=2 \pi n$}. Therefore, higher values of $n$ correspond to modes with higher UV contribution. When \mbox{$k \gg 1/ \ell $}, the factor $\cos{(k \ell)}$ in the components~\eqref{eq:cross} of the covariance matrix becomes highly oscillatory, leading to cancellations in the relevant integrals. Physically, UV modes correspond to short wavelengths in position space, which do not efficiently correlate spatially separated regions. As a result, we choose modes with small values of $n$---with sufficiently long wavelengths---in order to obtain stronger correlations. 

In addition, we consider that both regions contain the same number of modes, i.e.,~\mbox{$N_A=N_B=N$}, with~\mbox{$n=1,\dots, N$}. This is a reasonable choice because Eqs.~\eqref{eq:cross} depend on the product~$\tilde{g}^{(m)}( \black{k}) \tilde{g}^{(n)}(\black{k})$ (given explicitly by Eq.~\eqref{eq:fourier_transform}), and the dominant contributions to correlations arise when $n=m$. In fact, we verified that the more $n$ and $m$ differ, the less is their domain overlap, thereby becoming less relevant for correlations.

\subsection{Mode evolution}
With this setup, we now seek for a basis of solutions~$\varphi_\black{k}(t)$ to the time-dependent harmonic oscillator equation~\eqref{eq:harmonic}, and therefore a notion of vacuum. It is important to notice that the classical theory no longer has the full Poincaré group symmetry, specifically time-translational invariance. However, the electric field \eqref{eq:electric_pulse} vanishes asymptotically, thus recovering the Poincaré symmetry in both the asymptotic past and future. Hence, it is reasonable to work with a preferred basis of solutions denoted as the `in' solutions, whose behavior is the one of plane waves at $t \to - \infty$. They evolve non-trivially when the electric field is switched on and are globally given in terms of hypergeometric functions~\mbox{\cite{BeltranPalau2019, hypergeometric2025}}
\begin{align} 
        \varphi_\black{k}^\text{in}(t)\!=\!  \frac{e^{-i \omega^\text{in}_\black{k}t}}{\sqrt{2 \omega^\text{in}_\black{k}}}  \!\left(1+e^\frac{2t}{\Gamma} \right)^{\!\frac{1- i\delta}{2}}  \!\!\!
        {}_2F_1 \left(\rho_\black{k}^+, \rho_\black{k}^-,1\!-\! i \Gamma \omega_\black{k}^\text{in}; -e^\frac{2t}{\Gamma} \right), \nonumber\\[-3ex]
        {}\label{eq:in_solution}
\end{align}\par\vspace*{-2ex} \noindent
where
\begin{equation}
    \delta=\sqrt{(2 q E_0 \Gamma^2)^2-1},\quad \rho_\black{k}^\pm=\frac{1}{2} [1- i \Gamma (\omega_\black{k}^\text{in} \pm \omega_\black{k}^\text{out}) - i \delta].
    \label{eq:parameters_hypergeom}
\end{equation}
Here, $\omega_\black{k}^\text{in}$ and $\omega_\black{k}^\text{out}$ are the asymptotic values of the frequency $\omega_\black{k}(t)$ (see Eq. \eqref{eq:freq}) in the limits $t\to -\infty$ and $t \to +\infty$, respectively. These frequencies are time-independent since the vector potential $A(t)$ becomes constant in these limits. 

The dimensionless parameter~\mbox{$\delta=\sqrt{(2 qE_0 \Gamma^2)^2-1}$} plays a key role in the time evolution of the correlation measures, as it can be real or imaginary depending on whether $qE_0 \Gamma^2$ is larger or smaller than $1/2$, respectively. Consequently, different pulses yield qualitatively different trends: the particle production phenomenon is suppressed when~$qE_0 \Gamma^2 <1/2$, becoming relevant above that critical threshold. 

Finally, within the above setting, we fix $R/\Gamma=1$. Namely, we consider that the light crossing time of the regions where we localize the modes under study coincides with the effective duration of the pulse. Remember that, in order to have a sizeable amount of entanglement between the regions prior to the effect of the electric field, we had set $mR<1$. Now, choosing $R=\Gamma$ allows us to set the Schwinger ration $\varepsilon \equiv qE_0\Gamma^2/(m\Gamma)^2$ at relatively big values ($\varepsilon>1$), making sure that along the evolution the electric field becomes strong enough as to create a sensible amount of particles out of the vacuum. In other words, we make sure that we are exploring the non-perturbative regime of the Schwinger effect in which we are interested. 

\subsection{Correlations}
\begin{figure*} 
\centering
\begin{overpic}
[width=\linewidth]{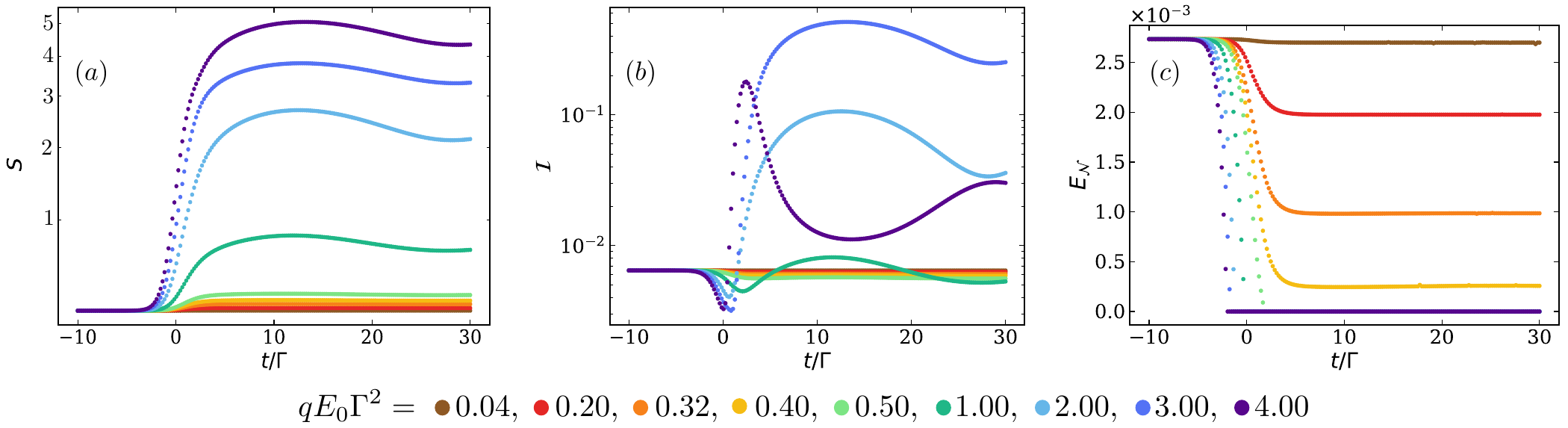}
\end{overpic}
\caption{
Time evolution of the von Neumann entropy $S$ (logarithmic scale), mutual information $\mathcal{I}$ (logarithmic scale), and logarithmic negativity $E_\mathcal{N}$ (linear scale) for different values of $qE_0 \Gamma^2$. Parameters: $m\Gamma=0{.}2$, $R=\Gamma$, $3 \times 3$ bipartition.
}
\label{fig:correlations_vs_t}
\end{figure*}

In Fig.~\ref{fig:correlations_vs_t},
we study the time evolution of the different correlation measures under the action of various electric pulses \eqref{eq:electric_pulse}, characterized by different values of $q E_0 \Gamma^2$. To this end, we have computed numerically the covariance matrix elements at each instant of time, from which the symplectic eigenvalues and, consequently, the correlation measures are obtained. We focus on the regime of electric strengths and pulse durations where the temporal switching efficiently triggers non-perturbative particle production, namely $\varepsilon>1$. In this regime, we have checked that decreasing the mass does not qualitatively alter the dynamic behavior of the correlation measures. Due to the computational cost of studying this time evolution, we have chosen a modest bipartition of $3\times 3$ modes. Despite the small number of modes per region, the obtained results are enough to qualitatively understand how the electric field affects the correlations between the modes across the regions.

In the initial stage \mbox{($t \lesssim -3\Gamma$)}, all curves for each measure converge to a common plateau, which corresponds to the value of each quantity when the electric field is not yet switched on. In contrast, the regions around~\mbox{$-\Gamma \lesssim t \lesssim +\Gamma$} present the most prominent changes since they coincide with the effective switch-on and -off of the electric field. As shown in Fig.~\ref{fig:correlations_vs_t}(a), increasing the value of the electric field implies an increase in the entropy. Higher values of $qE_0 \Gamma^2$ result in a sharper growth of this information measure, i.e., it reaches higher saturation levels over shorter timescales. Although the most significant growth occurs within the interval~\mbox{$-\Gamma \lesssim t \lesssim +\Gamma$}, the entropy continues to grow at a slower rate until well into the future, $t \approx 10 \Gamma$. 

Notably, for $t>0$ the electric field  starts to decrease, but the entropy has not yet stopped increasing. This reflects the non-Markovianity of particle production~\cite{Schmidt1999}: it is not a purely instantaneous effect of the field magnitude itself, but  depends on the integrated history of the pulse, determined by both the amplitude $E_0$ and the pulse duration~$\Gamma$. Once the electric pulse has effectively vanished and the entropy has peaked, the evolution is not stationary, as may be expected. Instead, the entropy exhibits an oscillatory trend. 

Let us analyze this long-time oscillatory behavior. Although the symplectic eigenvalues cannot be expressed in closed analytical form, an asymptotic expansion in the far-future limit can still be performed for the covariance matrix elements (see Eq.~\eqref{eq:cross}). For this purpose, consider the `out' modes in the limit $t \to +\infty$, which are described by their corresponding asymptotic plane wave behavior,
\begin{align}
    \varphi_\black{k}^\text{out} (t) \sim e^{-i \omega_\black{k}^\text{out}t}/{\sqrt{2 \omega_\black{k}^\text{out}}}.
\end{align}
Moreover, the asymptotic `in' modes can be expressed in terms of the `out' modes through the time-independent Bogoliubov coefficients $\alpha_\black{k}$ and $\beta_\black{k}$ as
\begin{align}
    \varphi_\black{k}^\text{in}  \sim \alpha_\black{k} \varphi_\black{k}^\text{out}+\beta_\black{k} (\varphi_\black{k}^{\text{out}})^*.
\end{align}
These coefficients must satisfy nontrivial constraints so that the Bogoliubov transformation preserves the canonical commutation relations, in particular
\begin{align}
    \abs{\alpha_\black{k}}^2- \abs{\beta_\black{k}}^2=1.
\end{align}

These   sets of modes ($\varphi_\black{k}^\text{in}$ and $\varphi_\black{k}^\text{out}$) define two different quantum vacuum states, the so-called `in' and `out' vacua, which may have relative excitations between them. In our  case,  relative excitations are a consequence of the Schwinger effect. The number of excitations is 
\begin{align}
    \mathcal{N}_\black{k}=\bra{\text{in}}  ((\hat{a}_\black{k}^{\dagger \text{out}} \hat{a}^\text{out}_\black{k}\!+\!\hat{b}_\black{k}^{\dagger \text{out}} \hat{b}_\black{k}^\text{out} )\ket{\text{in}},
\end{align} 
which, for the Sauter-type potential, admits the closed-form expression~\cite{BeltranPalau2019} 
\begin{equation}
    \mathcal{N}_\black{k}=2 \abs{\beta_\black{k}}^2=\frac{\cosh{[\pi (\omega_\black{k}^\text{out}- \omega_\black{k}^\text{in})\Gamma]} + \cosh{(\pi \delta)}}{\sinh{(\pi \omega_\black{k}^\text{in} \Gamma)} \sinh{(\pi \omega_\black{k}^\text{out} \Gamma)}},
    \label{eq:excitations}
\end{equation}
where the factor $2$ reflects the fact that particles and antiparticles are produced in pairs. 

Substituting the asymptotic behavior of the `in' solutions into the covariance matrix elements \eqref{eq:cross} reveals significant deviations from the vanishing electric field case. For instance, for large $t/\Gamma$, the `position' anticommutator can be decomposed into three distinct contributions as 
\begin{align} 
\label{eq:asymptotic_future}
        &\langle [\hat{Q}^m_I, \hat{Q}^n_J ]_+ \rangle (t) \sim \langle [\hat{Q}^m_I, \hat{Q}^n_J ]_+ \rangle_{E=0} \nonumber\\
        &\quad + \int \mathrm{d} \black{k}\, 
        \tilde{g}^{(m)}(\black{k}) \tilde{g}^{(n)}(\black{k})
        \frac{\mathcal{N}_\black{k}}{\omega_\black{k}^{\text{out}}}
        \cos(k \ell_{IJ}) \\
        &\quad + \int \mathrm{d} \black{k}\, 
        \tilde{g}^{(m)}(\black{k}) \tilde{g}^{(n)}(\black{k})
        \frac{2\,\mathrm{Re}\!\left(
        \alpha_\black{k} \beta_\black{k}^*
        e^{-2i \omega_\black{k}^{\text{out}} t}
        \right)}{\omega_\black{k}^{\text{out}}} \cos{(k \ell_{IJ})},
        \nonumber
    \end{align}
and likewise for the other components of the covariance matrix.
While the first term refers to the correlations before the electric field is switched on, the remaining two contributions encode the differences arising when the electric field is applied. The second term strongly depends on particle-antiparticle production via~$\mathcal{N}_\black{k}$, whereas the third is a time-oscillatory contribution.

Hence, the  non-stationary trend observed in Fig.~\ref{fig:correlations_vs_t}(a) can be attributed to the third term of the asymptotic expansion of Eq.~\eqref{eq:asymptotic_future} (and the similar ones appearing in the other covariance matrix components), thereby acting as a remnant interference contribution even after the external influence has ceased. Despite the non-stationary behavior observed in the asymptotic future, one can safely assert that the external classical electric field causes an overall increase in the entropy or, equivalently, in the mixedness of the reduced subsystems. This feature is fundamentally captured in the second term of the expansion \eqref{eq:asymptotic_future} (and analog ones), which is time-independent and is proportional to the density of excitations per mode, $\mathcal{N}_\black{k}$. As $\mathcal{N}_\black{k}$ grows, the probability of field modes being populated by created quanta increases. Hence, the external agent is effectively inducing a thermal-like noise into the local observables, thus contributing to a greater statistical mixing; that is, higher entropy values. As expected from the onset of the efficient Schwinger pair production, the strongest effects are observed when $qE_0 \Gamma^2 >1/2$, reflecting the relevance in that critical value for the dimensionless parameter $\delta$.

On the other hand, Fig.~\ref{fig:correlations_vs_t}(b) shows the evolution of the mutual information, which captures both classical and quantum correlations. We observe a non-trivial competition between the evolution of classical and quantum correlations. While Fig.~\ref{fig:correlations_vs_t}(a) clearly demonstrates that strong pulses imply a larger amount of classical mixedness, the entanglement decreases monotonically as shown in Fig. \ref{fig:correlations_vs_t}(c). Hence, it is reasonable to observe that decrease to a minimum near $t \approx 0$, when the quantum correlations decrease. However, classical correlations start to grow, thereby fighting against the dropping of the quantum ones, so the global result is the increasing of the mutual information. Once it reaches a maximum value, it starts to exhibit an oscillatory behavior, similar to the entropy in Fig. \ref{fig:correlations_vs_t}(a).

Lastly, the entanglement decay shown in Fig. \ref{fig:correlations_vs_t}(c) becomes more prominent when increasing $qE_0 \Gamma^2$. Specifically, for $qE_0 \Gamma^2 < 1/2$, once the electric field has been effectively switched off ($t/\Gamma \gg 1$), the behavior becomes stationary, as opposed to both the entropy and mutual information cases. In contrast, when the critical threshold is crossed ($q E_0 \Gamma^2  >1/2$), the logarithmic negativity vanishes completely at a finite time, rather than decreasing smoothly to a non-zero value. Furthermore, this vanishing point is not universal across cases; it occurs earlier as the parameter $qE_0 \Gamma^2$ increases. 

This can be understood in terms of the non-local nature of the entanglement. It is worth recalling that in Sec.~\ref{sec:setup} we defined an algebra of observables localized in two spatially disjoint regions. However, the Schwinger effect is based on the creation of excitations that are inherently delocalized, thereby spreading the initial entanglement between the corresponding modes in $A$ and $B$. This means that the classical electric field acts as a decohering agent, redistributing entanglement among other degrees of freedom rather than destroying it. 

It is well known that a noisy background, such as that generated by an external electric field, induces a decay of quantum correlations \cite{YuEberly2004}. Nevertheless, it is remarkable that over-critical pulses ($qE_0 \Gamma^2 >1/2$) lead to a vanishing logarithmic negativity. This does not necessarily imply that the modes are unentangled, since bound entanglement may still be present. This complete vanishing of the distillable entanglement at a finite time is analogous to the well-known phenomenon of entanglement sudden death \cite{YuEberly2004,YuEberly2009ESD}. In our context, however, this disentanglement is not caused by coupling to an external thermal reservoir, but rather by the non-perturbative excitation of the field modes driven by the classical background, which redistributes the initial vacuum correlations across delocalized degrees of freedom.

\begin{figure} \centering
\includegraphics[width=0.95\columnwidth]{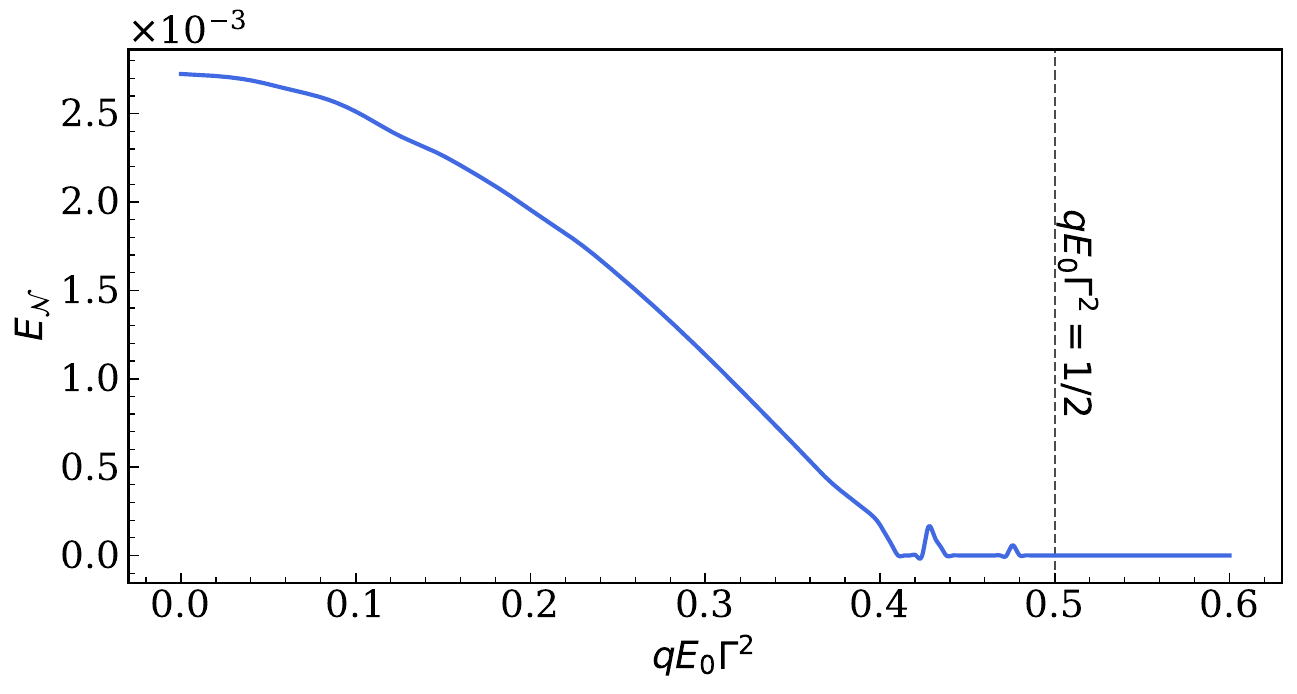}
	\caption{Transition of the spontaneous entanglement at~\mbox{$t/\Gamma=25 \gg1$}. Parameters: $m\Gamma=0{.}2$, $R=\Gamma$, $3 \times 3$ bipartition.}
	\label{Fig:phase_transition}
\end{figure}

Additionally, the transition from a non-vanishing logarithmic negativity when the electric field is switched off to a vanishing one suggests the presence of a phase transition, in which $qE_0 \Gamma^2$ and $E_\mathcal{N}$ represent the control and order parameter, respectively. Fig. \ref{Fig:phase_transition} shows the logarithmic negativity evaluated in the asymptotic future for different values of $q E_0 \Gamma^2$. It decreases until $q E_0 \Gamma^2 \approx 0{.}4$, where a slight oscillatory trend is observed, before vanishing around the critical point $qE_0 \Gamma^2\approx 1/2$. The onset of the transition below this value reflects the fact that the dimensionless parameter $qE_0 \Gamma^2$ may not be the only relevant parameter governing it, although it plays a dominant role. Besides, the small number of modes per region considered might be responsible for finite-size scaling effects. Further detailed analysis of this potential phase transition will be carried out elsewhere. 

\section{Correlations under electric fields in $1+3$ dimensions}\label{sec:electric_3d}

In three spatial dimensions the analysis of correlations under electric fields introduces a new feature: a preferred direction in the system (that of the electric field), and the question arises how this anisotropy affects quantum correlations across different regions.

The simplest setting to explore this question is the generalization to three dimensions of the one-dimensional setting studied previously. Namely, to consider that regions $A$ and $B$ are two spherical balls of radius $R$ separated a distance $\ell =2R$, so that they are adjacent. However, within this setting,  the small number of modes that our computational capabilities allows us to prove, is not enough as to capture a non-negligible amount of distillable entanglement, even prior to the effect of the electric field (in our simulations we have always obtained a vanishing logarithmic negativity when the electric field vanishes). 

This fact is easy to understand. Not only in three dimensions entanglement is more diluted because it has many more ways to spread, but also the single point shared by the region's boundaries is not enough to retain a decent amount of entanglement across regions for a reasonable number of modes. Increasing the number of modes supported in each region would favour the presence of correlations, but this is still not an efficient method to obtain a sensible amount of quantum correlations across these regions, because of the spatial separation between their boundaries. 

Therefore, a detailed study of correlations under the effect of an electric
field in $(1+3)$ dimensions requires a configuration of regions and modes with
the following properties. First, we should define such configuration  in the asymptotic past, where there is no electric field, and follow it throughout the duration of the electric
pulse. Second, it should be sensitive to anisotropic effects. Finally,  the relevant modes
should  be constructed at low numerical cost along the evolution.

Previous studies in pure Minkowski spacetime (without electric fields) have obtained non-vanishing logarithmic
negativity in two different settings~\cite{perche2026, ubiquitousQFT}.
Ref.~\cite{perche2026} considers that regions $A$ and $B$ are disjoint spherical
balls (like our setting), and finds the shape of the single-modes in each region
containing most of the entanglement between them. These  ``most entangled
modes'' are obtained by taking linear combinations of more than one thousand
cylindrically symmetric modes in each region, which makes the construction
costly to repeat at every time step of the evolution. Ref.~\cite{ubiquitousQFT}
considers concentric spherical regions (spherical layers) in order to obtain a
non-vanishing logarithmic negativity among some modes supported across those
regions. Being spherically symmetric, however, this configuration cannot probe
anisotropic effects. Identifying configurations that combine maximal entanglement with anisotropic sensitivity is an interesting direction that we leave for future work.

For this reason, we consider two adjacent balls and a small number of modes. This setting is simple to
construct along the evolution and captures the anisotropy of the field
correlations, at least in those encoded in the mutual information.

Let us then study mutual information in $(1+3)$ dimensions closely following the analysis of Sec.~\ref{sec:electric_1d}, with the following additional features. We choose the orthonormal set of spherically symmetric smearing functions and their corresponding Fourier transform:
\begin{equation}
    g^{(n)}(\black{x})\!=\!\frac{\sin{(2 \pi n r)}}{\sqrt{2 \pi}r} \Theta(1-r),\quad 
    \tilde{g}^{(n)}(\black{k})\!=\! \frac{2 n R^2\sin(kR)}{k[(kR)^2\!-\!4 \pi^2 n^2]} ,
    \label{fourier_3d}
\end{equation}
where $ r=\abs{\black{x}}/R$ and $k=\abs{\black{k}}$. Note that, as happened in $(1+1)$ dimensions, $g^{(n)}(\black{x})$ becomes more oscillatory as $n$ increases or, equivalently in momentum space, $\tilde{g}^{(n)}(\black{k})$ is mostly localized around $kR=2 \pi n$. Hence, the behavior of the correlations when changing the main parameters, such as the number of modes under consideration, is expected to be qualitatively similar to the $1+1$ case, although with a lower value, since the distribution of correlations becomes more diluted.

The anisotropy is controlled by the angle $\gamma$ between the electric pulse $\black{E}(t)$ and the vector $\bm{\ell}$ joining the centres of the balls. The covariance matrix elements~\eqref{eq:cross} are now calculated with a triple integral in which the corresponding azimuthal integral can be performed analytically. In order to illustrate it, consider the example of the position anticommutator:
\begin{align}
&\langle [ \hat{Q}^m_I, \hat{Q}^n_J ]_+\rangle(t)
\!=\! 4\pi\!\! \int\!\! \mathrm{d}k \, \mathrm{d}\theta  k^2\! \sin\theta 
\tilde{g}^{(m)}(\mathbf{k}) \tilde{g}^{(n)}(\mathbf{k})
\!\left|\varphi_{\mathbf{k}}(t)\right|^2 \nonumber\\
& \qquad\quad\times \cos(k \ell_{IJ}\cos\theta\cos\gamma)\,
J_0(k \ell_{IJ}\sin\theta\sin\gamma),
\end{align}
where $J_0$ denotes the Bessel function of the first kind and zero order. The other two integrals (the radial $k$ and the polar $\theta$ ones) have to be performed numerically.

\begin{figure} \centering
\includegraphics[width=0.95\columnwidth,height=0.65\columnwidth]{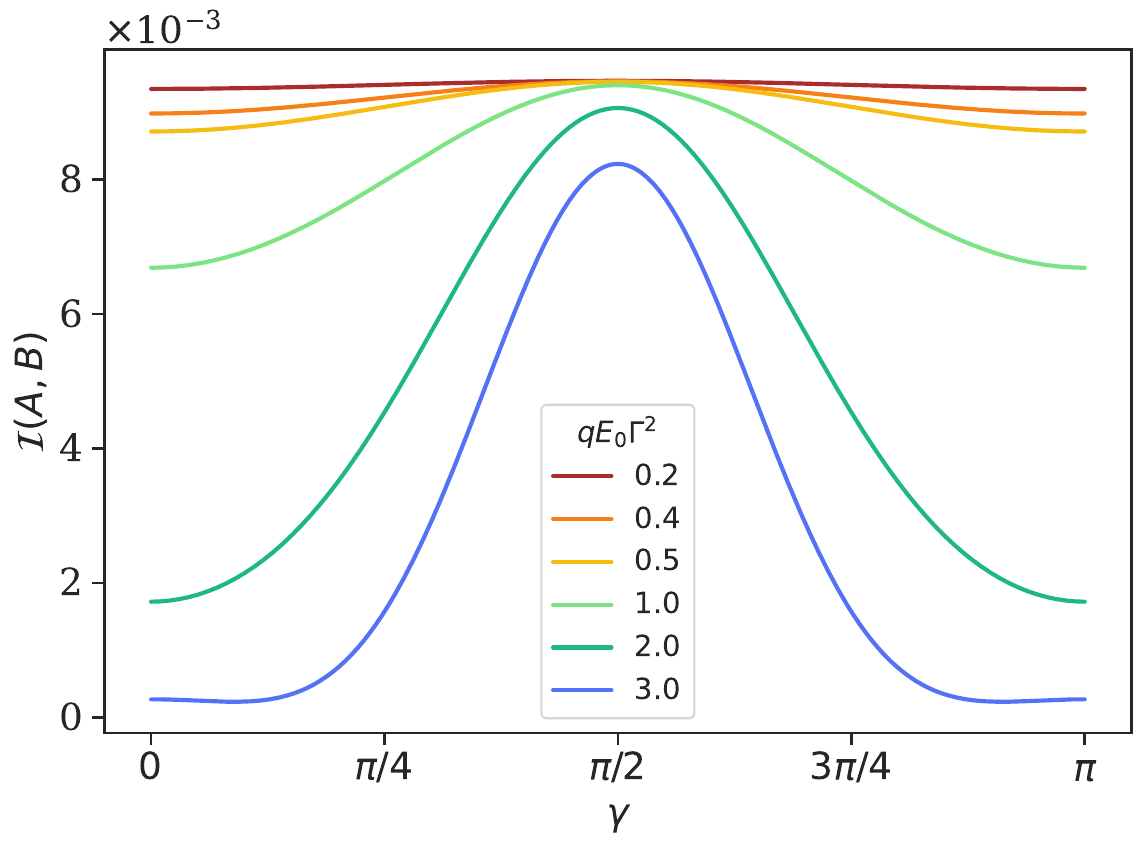}
	\caption{Mutual information anisotropy at $t=0$ for different values of~$qE_0 \Gamma^2$. Parameters: $m\Gamma=0{.}2$, $R=\Gamma$, $3 \times 3$ bipartition.}
	\label{Fig:anisotropy}
\end{figure}

This anisotropic behavior is illustrated in Fig.~\ref{Fig:anisotropy}, where
the mutual information is evaluated at $t=0$ (corresponding to the peak of the electric pulse) for different orientations of the regions. This analysis is repeated for several values of $qE_0 \Gamma^2$. All curves are symmetric about $\gamma=\pi/2$, at which the mutual information attains its maximum value. Therefore, the selected modes become more correlated when the electric field is orthogonal to the line connecting the regions. Moreover, this anisotropic enhancement becomes increasingly pronounced as the pulse intensity grows. This makes sense because an orthogonal electric field contributes less efficiently to the mixedness among the regions than a parallel one, thus decreasing the joint entropy $S(AB)$ and, in turn, increasing the mutual information.

\section{Conclusions}\label{sec:conclusions}

In this work, we have studied the distribution of classical and quantum correlations in the vacuum state of a charged scalar field in the presence of a homogeneous and time-dependent electric field (in both $(1+1)$ and~$(1+3)$ dimensions). We constructed a finite set of localized Gaussian observables by smearing the field operators within two adjacent spatial regions of spherical symmetry, thereby enabling the application of QI techniques in a QFT framework. This approach extends previous studies that focused primarly either on correlations between momentum modes \cite{Kranas2023}, or on real-space correlations in static situations \cite{ubiquitousQFT}. 

We then considered a Sauter-type electric pulse and observed that the vacuum correlations evolve non-trivially in time. Specifically, the electric field enhances the entropy of the regions, thus revealing a larger degree of mixedness, while the mutual information typically grows as well. Moreover, considering intense pulses leads to more significant variations, therefore reflecting the onset of the efficient Schwinger pair production when the critical threshold is exceeded ($qE_0 \Gamma^2>1/2$). In contrast, the logarithmic negativity decreases more rapidly with strong pulses as the pair production redistributes the initial entanglement among additional degrees of freedom. For sufficiently strong pulses in the non-perturbative regime, the logarithmic negativity vanishes at a finite time, a behavior consistent with the phenomenon known as entanglement sudden death. In the asymptotic future, this manifests as a threshold-like behavior predominantly governed by $qE_0\Gamma^2$. A detailed characterisation of this potential phase transition and its full parameter space is left for future work. 

Finally, extending the analysis to $(1+3)$ dimensions allowed us to study the anisotropy induced by the electric field. This extends previous studies of particle production, which have predominantly focused on isotropic backgrounds, such as FLRW cosmologies~\cite{Parker1968, Parker1969, Parker1971}.  In particular, studying, at a fixed time, the dependence of the correlations on the relative orientation between the localized regions and the electric field revealed significant variations in the mutual information, with the maximum occurring when the line joining the centres of the regions is orthogonal to the field direction. These results show that the external agent not only modifies the amount of correlations but also their spatial distribution. \\

\begin{acknowledgments}
The authors would like to thank Patricia Ribes-Metidieri, Javier Olmedo and Carlos Barceló for helpful discussions. AAD, LJG and MMB acknowledge  support through the Grant PID2023-149018NB-C44 (funded by MICIU/AEI/10.13039/501100011033). This work was also made possible through the support of the WOST, WithOut SpaceTime project (https://withoutspacetime.org), supported by Grant ID\# 63683 from the John Templeton Foundation (JTF). The opinions expressed in this work are those of the authors and do not necessarily reflect the views of the John Templeton Foundation. The work of Iker Sanz-González is partially supported by Spanish
MINECO/FEDER Grants
PID2024-160228NB-I00, funded by MCIN/AEI/10.13039/501100011033 and
“ERDF A way of making Europe”.
\end{acknowledgments}

\bibliographystyle{jhep-en}
\bibliography{references}

\end{document}